\documentclass[conference]{IEEEtran}
\IEEEoverridecommandlockouts
\usepackage[nocompress]{cite}
\usepackage{multicol}
\usepackage{amsmath,amssymb,cite,multirow,subfigure}
\usepackage{graphicx}
\usepackage{color}
\usepackage[linesnumbered,ruled, vlined]{algorithm2e}
\usepackage{makecell}
\usepackage{threeparttable}
\usepackage[usenames,dvipsnames,svgnames,table]{xcolor}
\usepackage{enumerate}
\usepackage{amsmath,amsthm}

\usepackage{multicol}
\usepackage{multirow}
\usepackage{bm,amsmath}

\newcommand{\tabincell}[2]{\begin{tabular}{@{}#1@{}}#2\end{tabular}}

\usepackage{algorithmic}

\usepackage[misc]{ifsym}

\usepackage{verbatim}
\usepackage{url}
\usepackage{makecell}

\usepackage{booktabs}
\usepackage{threeparttable}

\usepackage{ulem}
\usepackage{soul}
\newcommand{\RF}[1]{\textcolor{black}{#1}}
\newcommand{\MX}[1]{\textcolor{black}{#1}}
\newcommand{\WX}[1]{\textcolor{black}{#1}}
\newcommand{\F}[1]{\textcolor{black}{#1}}

\makeatletter
\renewcommand\normalsize{%
\@setfontsize\normalsize\@xpt\@xiipt
\abovedisplayskip 5\p@ \@plus2\p@ \@minus5\p@
\abovedisplayshortskip \z@ \@plus3\p@
\belowdisplayshortskip 6\p@ \@plus3\p@ \@minus3\p@
\belowdisplayskip \abovedisplayskip
\let\@listi\@listI}
\makeatother

\makeatletter

\newcommand{\Rmnum}[1]{\expandafter\@slowromancap\romannumeral #1@}
\makeatother

\usepackage{stfloats}
\fnbelowfloat

\def\BibTeX{{\rm B\kern-.05em{\sc i\kern-.025em b}\kern-.08em
    T\kern-.1667em\lower.7ex\hbox{E}\kern-.125emX}}
\begin{document}

\title{An FPGA-Based Accelerator Enabling Efficient Support for CNNs with Arbitrary Kernel Sizes
\vspace{-0.1in}}

\vspace{-1in} 
\author{

    Miaoxin~Wang$^1$, Xiao~Wu$^1$, Jun~Lin$^1$$^{(\textrm{\Letter})}$, and Zhongfeng~Wang$^{1,2}$$^{(\textrm{\Letter})}$\\
	\IEEEauthorblockA{
		$^1$ School of Electronic Science and Engineering, 
		Nanjing University, Nanjing, China\\
            $^2$ School of Integrated Circuits, Sun Yat-sen University, Shenzhen, China\\
		Email:\{mxwang, wxiao\}@smail.nju.edu.cn,
          \{jlin, zfwang\}@nju.edu.cn
    }
    \vspace{-0.5in}
}



\maketitle
\vspace{-1in}

\begin{abstract}
Convolutional neural networks (CNNs) with large kernels, drawing inspiration from the key operations of vision transformers (ViTs), have demonstrated impressive performance in various vision-based applications. 
To address the issue of computational efficiency degradation in existing designs for supporting large-kernel convolutions, an FPGA-based inference accelerator is proposed for the efficient deployment of CNNs with arbitrary kernel sizes. 
Firstly, a Z-flow method is presented to optimize the computing data flow by maximizing data reuse opportunity. 
Besides, the proposed design, incorporating the kernel-segmentation (Kseg) scheme, enables extended support for large-kernel convolutions, significantly reducing the storage requirements for overlapped data. 
Moreover, based on the analysis of typical block structures in emerging CNNs, vertical-fused (VF) and horizontal-fused (HF) methods are developed to optimize CNN deployments from both computation and transmission perspectives. 
The proposed hardware accelerator, evaluated on Intel Arria 10 FPGA, achieves up to \F{3.91$\times$} better DSP efficiency than prior art on the same network. 
Particularly, it demonstrates efficient support for large-kernel CNNs, achieving throughputs of 169.68 GOPS and 244.55 GOPS for RepLKNet-31 and PyConvResNet-50, respectively, both of which are implemented on hardware for the first time.

\end{abstract}


\vspace{-0.12in}
\section{Introduction}
\vspace{-0.0395in}
CNNs exhibit prominent performance in many fields of computer vision.
Recently, they have been strongly challenged by ViTs \cite{Vit2021}, which have gained widespread recognition with unexpected success by utilizing self-attention mechanisms.
\WX{Inspired by the ViTs' key operation, some remarkable works introduce convolution operations with large kernel sizes to expand the receptive field of CNNs for pursuing higher accuracy.}
For example, RepLKNet\cite{RepLKNet2022} with kernel size ranging from 3$\times$3 to 31$\times$31, has achieved comparable or even superior results to Swin Transformer in multiple downstream tasks.

The variety of kernel size among different layers poses a significant challenge for the hardware implementation of CNNs, \WX{since previous works are} typically tailored for the deployment of CNNs with stacks of small convolutions (e.g., 3$\times$3).
Directly supporting large kernels with a general convolution engine can result in significant storage overheads, \WX{as the duplicate data caused by window slippage are typically buffered in on-chip memory for subsequent computations \cite{MA2018TVLSI}.}
Recent works have attempted to address this issue by dividing large convolution kernels into several smaller sub-kernels.
Unfortunately, due to their fixed hardware architecture, these designs only support convolutions with certain sizes \cite{split2021TCAS2-1} or simply insert zero elements to ensure workload balance \cite{split2021TCAS2-2}, which results in unacceptable computational inefficiencies.
\WX{Besides, directly decomposing large-kernel operations without consideration of data arrangement in hardware implementation can cause a significant decrease of accuracy \cite{split2018ISVLSI}.}

Moreover, \WX{these large-kernel operations are widely used in many typical block structures, such as  MBconv block in light-weight CNNs\cite{EfficientNet2019} and the block based on multi-scale convolutions in \MX{NAS-based} CNNs \cite{pyconv2022TEC}.}
\WX{However, most existing works process the emerging CNNs based on a layer-by-layer execution scheduling, ignoring the potential optimization opportunities of the repeated block structures.}
\WX{Although some works attempt to optimize the layer execution scheduling through fusion methods, their fixed hardware architecture hinders them from efficiently supporting different types of layers \cite {2021FPL},\cite{MobileNetV2_1}.
Besides, these approaches often require additional on-chip resources to buffer the intermediate results of fused layers, resulting in substantial storage consumption \cite{MobileNetV2_1}.}
To address the above issues, a flexible FPGA-based hardware accelerator is proposed, the main contributions of which are highlighted as follows.
\begin{itemize}
\item \textbf{Layer-level optimization:} \WX{An optimized dataflow collaborating with Z-flow method and Kseg scheme is developed for convolution operations of arbitrary kernel sizes.}
\item \textbf{Block-level optimization:} Based on the detailed analysis of typical block structures, \WX{a flexible layer execution strategy integrating VF and HF methods is developed to further improve the overall throughput.}
\item \textbf{Experiment Results:} \WX{The proposed hardware accelerator is implemented on Intel Arria 10, achieving up to \F{3.91$\times$} better DSP efficiency compared to existing FPGA accelerators evaluated on the same network.}

\end{itemize}
\vspace{-0.1in}
\section{Proposed hardware architecture}
\vspace{-0.02in}
\vspace{-0.04in}
\subsection{Architecture Overview}
\vspace{-0.04in}

\begin{figure*}[t]
\vspace{-1em}
  \centering
  \includegraphics[width=0.99\textwidth,height=0.38\textwidth]{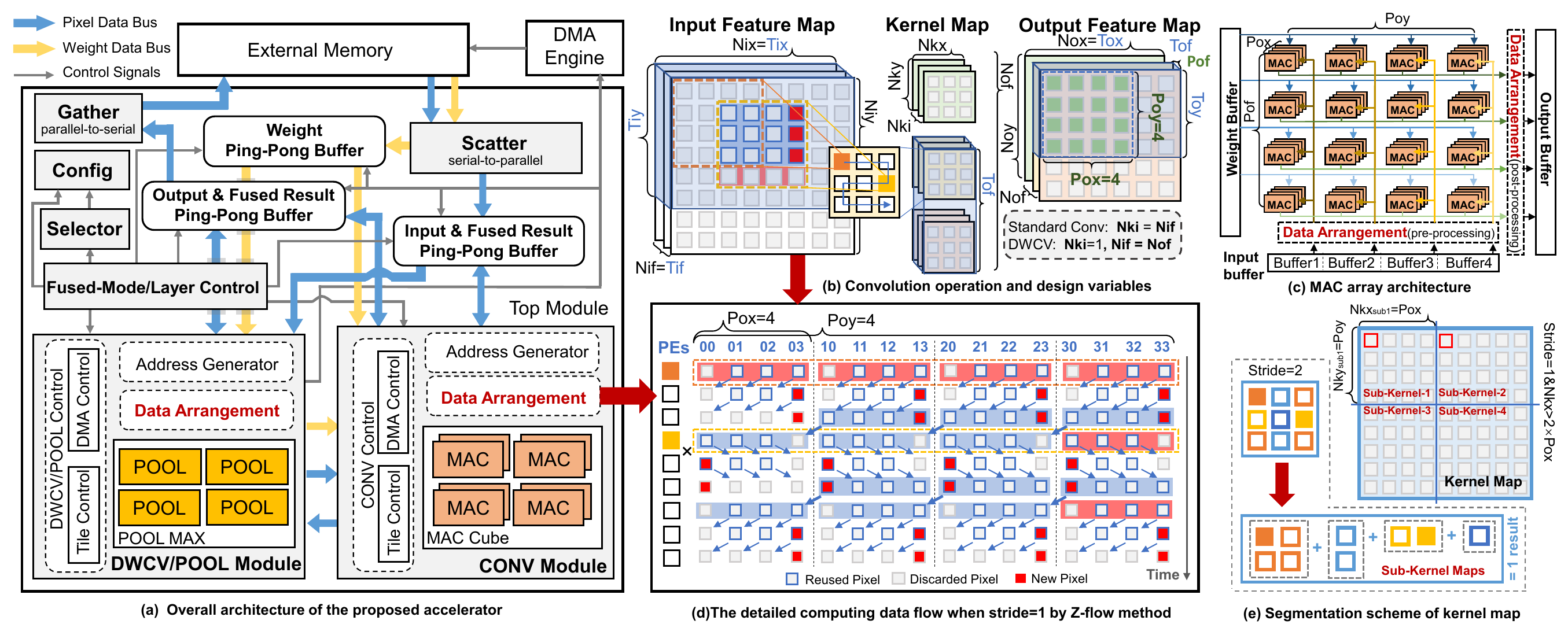}
  \vspace{-0.14in}
  \caption{The overall architecture and computing dataflow. (a) The overall architecture of the proposed accelerator. (b) Convolution operation and design variables. \F{(c) MAC array architecture. }(d) Detailed computing dataflow of the Z-flow method when the stride is 1. \F{(e) Segmentation scheme of kernel map. }
  }\label{Fig1}
\vspace{-0.25in}
\end{figure*}


\WX{The overall architecture of the proposed accelerator is illustrated in Fig.~\ref{Fig1} (a), which can be mainly divided into three parts.}
In the \textbf{storage system}, the direct memory access (DMA) engine fetches data from \WX{external memory}, while the on-chip buffers are responsible for buffering tiles of the feature and kernel maps.
We set $Tix=Nix$ and $Tif=Nif$ to generate complete results for an entire row, \WX{where the loop tiling variables $T*$ determine the sizes of buffered data and loop dimension variables $N*$ denote the sizes of the feature and kernel maps (in Fig.~\ref{Fig1} (b)).}
\WX{In the \textbf{control system}, the fused- mode/layer control module allocates control signals to ensure coordinated and efficient execution of given CNNs based on different layer execution scheduling.}
\WX{The parameters required for the computation of each layer are pre-stored in configuration registers.}
\WX{The \textbf{computing system}, equipped with a processing element (PE) array consisting of $Pox \times Poy \times Pof$ multiply-accumulate (MAC) units (in Fig.~\ref{Fig1} (c)), is specifically designed for standard convolution (CONV) and depthwise convolution (DWCV) with arbitrary kernel sizes.}
\WX{The parallel computations, determined by the loop unrolling variables $P*$, are employed along dimensions of $Pox$ and $Poy$ within one input feature map and across $Pof$ kernel maps.}
\vspace{-0.06in}
\subsection{Computing Dataflow for Arbitrary-Kernel CONV (AKCV)}
\vspace{-0.02in}
\WX{Previous works \cite{split2021TCAS2-1} \cite{split2021TCAS2-2}, considering the constraint imposed by their fixed hardware structure, simply split the large kernel maps into smaller ones for adaptation.}
When the kernel size cannot be evenly divided, these small kernels are padded with zero elements for a unified process, which causes significant redundant computations.
Moreover, line buffers are adopted in many works\cite{2021FPL}\cite {Chao2021iscas} to store sliding overlapped pixels.
The size of these buffers is directly proportional to the kernel size, which restricts their capability to support convolutions with larger kernels.
Thus, to enhance the efficiency for supporting AKCV operations, a flexible dataflow collaborating with the Z-flow method and Kseg scheme is proposed, which significantly reduces on-chip memory consumption by fully utilizing the data reuse opportunities.
The specific descriptions of these optimization methods are provided as follows.

\subsubsection{\textbf{Z-Flow method}}
\WX{Since the dimensions of the kernel map aren't processed in parallel based on our loop unrolling strategy, the multiplications of weight and its corresponding inputs are performed in chronological order, as shown in Fig.~\ref{Fig1} (d).}
The detailed data arrangement of the proposed Z-Flow method can be discussed in the following two parts.
\WX{Firstly, we take the computing dataflow in PE$\_0i$ $(0i=00,01,02,03)$ for the first three cycles as an example to explain how to utilize the data reuse opportunities offered by the X-dimension.}
In the beginning, four pixels from in- put buffers are loaded into registers and then sent to the corresponding PE$\_0i$. 
These pixels are reused by the left adjacent register array in the next cycle, while the rightmost one in each register array starts reading input pixels from buffers, which are denoted in red in Fig.~\ref{Fig1} (b) and (d).
In the subsequent cycles, repeat this operation until the computation of one row of data in the kernel map is completed.
When computing the even row in kernel maps, the pixels perform the mirroring operations, reused by the right adjacent register array.
\WX{Secondly, the proposed Z-flow method utilizes the data reuse opportunities of Y-dimension provided by the inflection point in the kernel maps, as highlighted in yellow in Fig.~\ref{Fig1} (b) and (d).}
Specifically, in the fourth cycle, the pixels that were sent to PE$\_1i$ last cycle are reused by the left adjacent register arrays, namely sent to corresponding PE$\_0i$. 
Meanwhile, the new pixels are fed into the rightmost register arrays from the input buffers.
Additionally, a small number of data required for the subsequent computations are prefetched and stored in the combination registers to avoid reading conflicts.

\subsubsection{\textbf{Kernel-segmentation (Kseg) scheme}}
The larger kernel maps ($Nkx$\textgreater $2\times Pox$) need to be segmented in our design to avoid extra resource consumption.
We set the size of the first sub-kernel as an integer multiple of the loop unrolling variables, e.g., $Nkx_{sub1} = Pox$ in Fig. \ref{Fig1} \F{(e)}, to ensure that the corresponding input pixels required for the computations of other sub-kernels can be exactly obtained from the same address of different parallel buffers without read conflict.
When processing the computation with a stride size of 2, the Kseg scheme based on the mapping of input and weight is shown on the left of Fig. \ref{Fig1} \F{(e)}.
\WX{By this means, different sub-kernels with varying kernel sizes can be executed directly by the proposed Z-flow method, without introducing redundant computations.}
The results obtained from these sub-kernels are accumulated to generate the final outputs.
\begin{figure}[t]
\vspace{-0.1in}
  \centering
  \includegraphics[width=8.70cm]{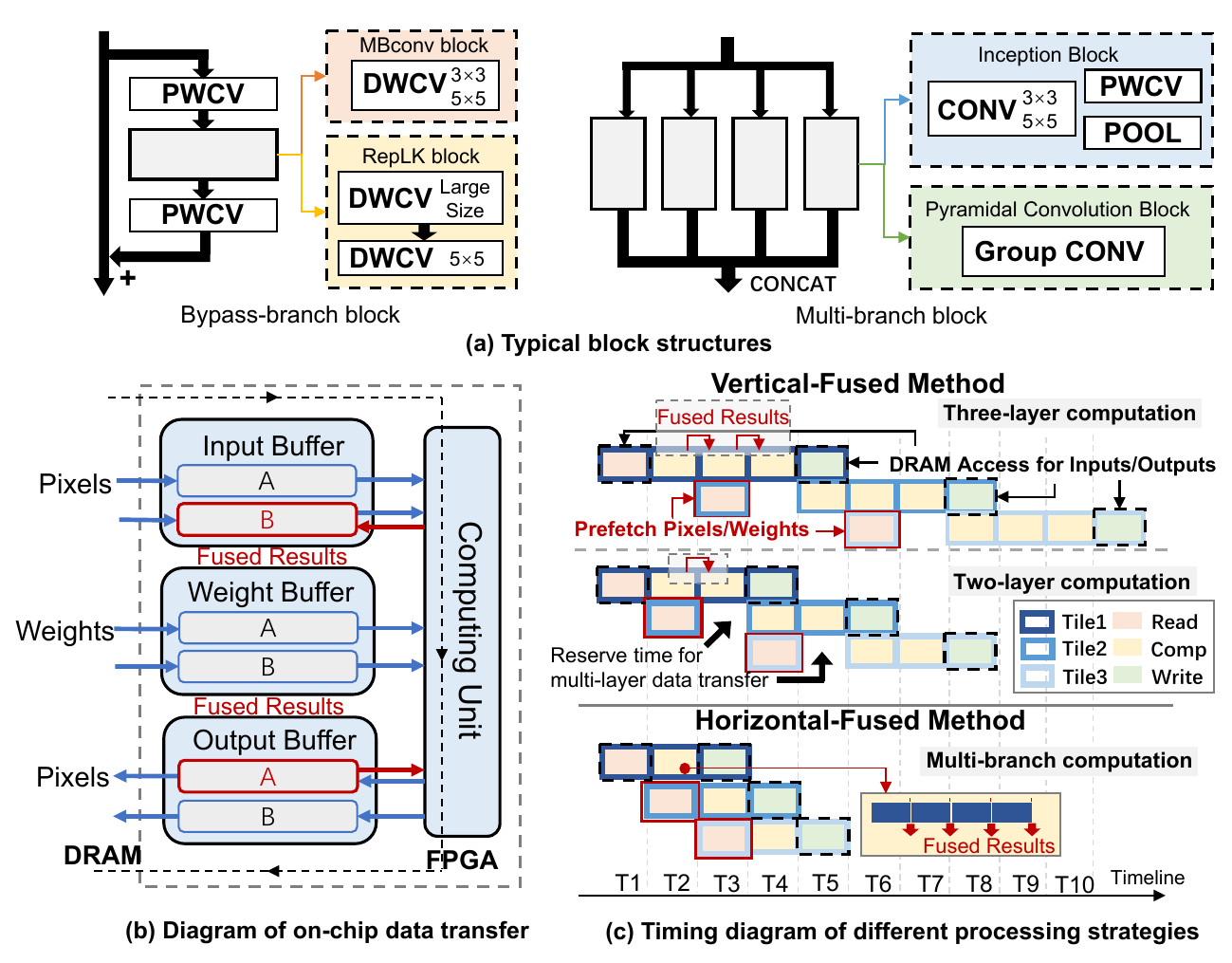}
  \vspace{-0.1in}
  \caption{Fusion methods and \RF{corresponding execution scheduling}. (a) Structures of typical blocks. (b) Data transactions between on-chip buffers and computing units. (c) \RF{The execution scheduling} of different fusion strategies.
  }\label{Fig2}
  \vspace{-0.28in}
\end{figure}
\vspace{-0.11in}
\subsection{Flexible Layer Execution Strategy}
\vspace{-0.05in}
\RF{Common block structures can be broadly divided into two categories: bypass-branch block and multi-branch block, as depicted in Fig.~\ref{Fig2} (a).}
For the former, one layer in this block relies on the outputs of its previous layer for computation, whereas for the latter, different branches are processed independently.
Based on the characteristics of different structures, we propose a layer execution strategy, incorporating VF and HF methods, to efficiently support these blocks.
\RF{Specifically, different from previous approaches that require additional on-chip storage to buffer intermediate fused results, our VF method makes full use of existing resources to eliminate redundant off-chip accesses, as shown in Fig.~\ref{Fig2} (b).}
\RF{Through the HF method, the processing of the multi-branch block is optimized by reducing redundant data transfers and balancing workloads among PEs.}
\RF{The detailed execution scheduling of the above methods is exhibited in Fig.~\ref{Fig2} (c).}

\subsubsection{\textbf{VF Method}}
The VF method manages data transactions between on-chip buffers and computing units to store intermediate fused results without additional storage overhead.
By this means, the computation of fused layers can be performed continuously, greatly enhancing computational efficiency.
\WX{To simplify the control logic complexity, off-chip read/write operations for pixels are still performed by the respective input/output buffers interacting with dynamic random access memory (DRAM).}
\WX{Moreover, thanks to the loop tiling strategy and dual buffer technique, fine-grained optimization is achieved through the overlapping of input data prefetched for the next tile with computation operations.}
\WX{The detailed tile-by-tile execution scheduling of the MBconv block is taken as an example for further discussion, considering both data transfer and computation aspects.}
\WX{Benefiting from the proposed VF method, only the input pixels of the first-fused layer, the weights of all-fused layers, and the output pixels of the third-fused layer need to be transferred.}
\WX{The specific block-level optimization between computation and transmission can be discussed in the following three cases.}
Firstly, for the first tile, since data transactions for the computation of the first-fused layer involve input buffer A and B, reading data from DRAM for the next tile (Trans\_next) cannot be performed until completing this computation to avoid read-write conflicts.
Secondly, for the last tile, since output buffer A and B are used for data transfer in the third-fused layer, writing data to DRAM for the previous tile (Trans\_prev) needs to be completed before this computation. 
Thirdly, for the other tiles, Trans\_prev is executed first to ensure that Trans\_prev and Trans\_next can overlap with the computation of all-fused layers without the above conflicts.
\subsubsection{\textbf{HF Method}}
For multi-branch block, considering parallel layers share the same inputs, redundant data transfers can be avoided in the HF method by transferring inputs only once.
\WX{Moreover, some multi-branch structures in emerging models often incorporate multi-scale kernels to increase the receptive field for higher accuracy, whereas the mismatch between model size and computation parallelism may cause significant computational inefficiency.}
\WX{To elaborate on the computation optimization, we take the PyConv block \cite{pyconv2020arXiv} as an example.}
\WX{In this block, group CONV is employed to ensure even distributions of computational cost among different branches, with the number of output channels per group ($Nof\_group$) decreasing as the kernel size increases.}
\WX{However, this characteristic results in a mismatch between $Nof\_group$ and $Pof$, significantly impacting PE utilization, with the worst-case scenario being 1/16.}
The HF method alleviates this problem by executing different group CONV in parallel.
To simplify the dataflow, kernels from different branches are unified to the same size ($Nkx\_unify$) without introducing additional computation latency.
Thus, the computation latency optimized by the HF method ($T_{comp\_HF}$) can be expressed as
\begin{equation}\label{eq2.1}
\begin{small}
\begin{aligned}
& T_{comp\_HF} = 
Group\_num \times Nkx\_unify \times Nky\_unify\\
&  \times Nif\_group \times \frac{Nox \times Noy \times sum(Nof\_group\_i)}{Pox \times Poy \times \F{Pof}}\times clk\times 10^3. &
\end{aligned}
\end{small}
\end{equation}
where $Group\_num$ means the number of groups and $\_i$ means the parameter of the $i$-$th$ parallel layer.
The output pixels of these parallel layers, continuously generated at different times based on their group numbers, are rearranged and sent to the corresponding address in output buffers.

\vspace{-0.04in}
\section{Experimental results}
\subsection{Experimental Settings}
\vspace{-0.02in}
The proposed accelerator is evaluated on Intel Arria 10 SoC FPGA, running at the frequency of 200MHz, with \F{55}\% ALM, 30\% DSP, and 48\% RAM used, where the resource usage is determined by the loop design variables, e.g., $T*$ and $P*$.
We set $Pox, Poy$, and $Pof$ as $8, 8, 16$, to configure the size of the computing engine.
An 8-bit data quantization scheme is adopted in our design, which is consistent with the scheme used in the works selected for comparisons.
\vspace{-0.08in}
\subsection{FPGA Implementation Results}
\vspace{-0.02in}
\begin{figure}[t]
\vspace{-0.1in}
  \centering
  \includegraphics[width=8.70cm]{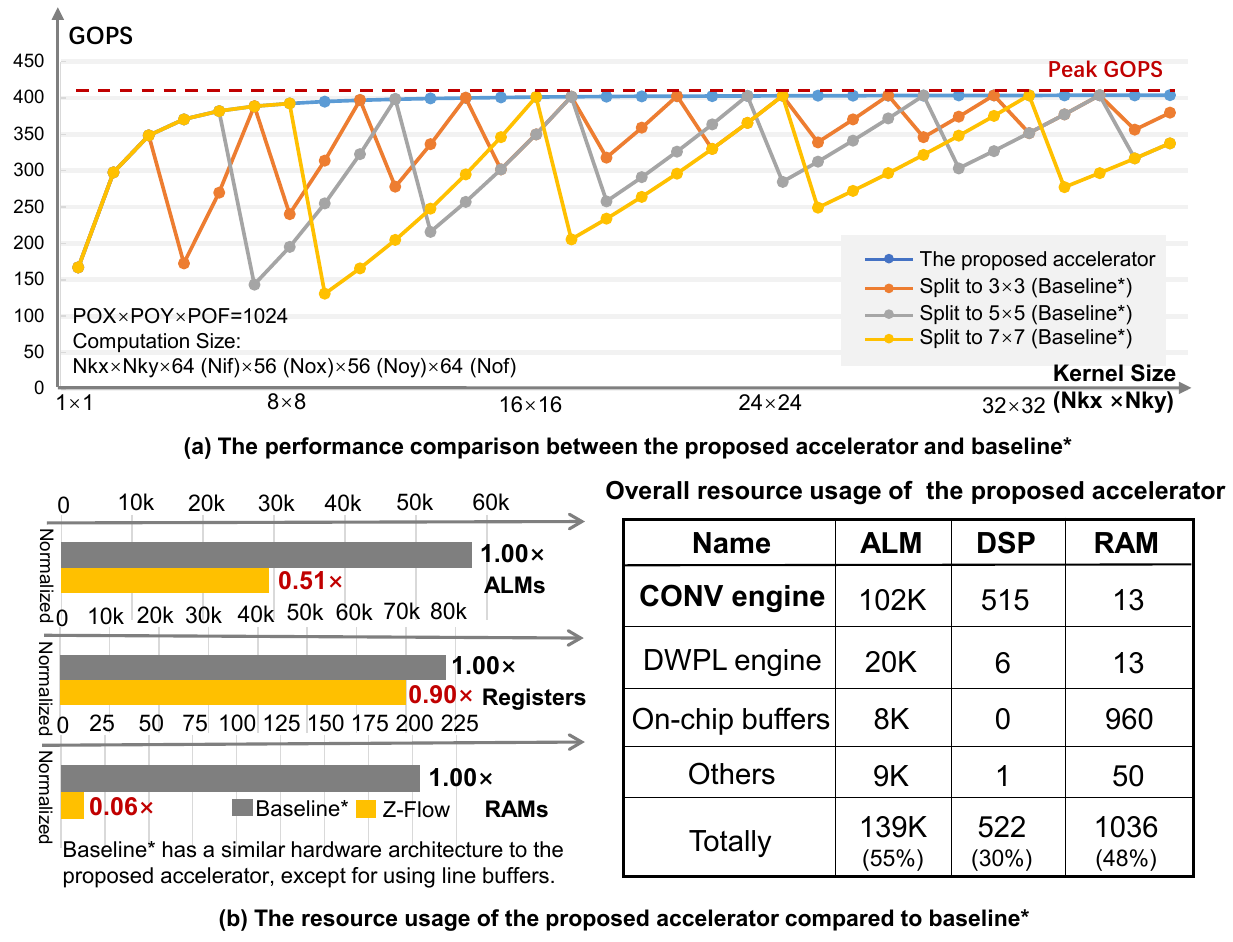}
  \vspace{-0.1in}
  \caption{Performance results and storage resource \WX{usage} of layer-level optimization. (a) The comparison of throughput under different kernel sizes. (b) The reduction of resource usage for data arrangement in the CONV engine.
  }\label{Fig3}
 \vspace{-0.25in}
\end{figure}
\begin{figure}[t]
\vspace{-0.1in}
  \centering
  \includegraphics[width=8.70cm]{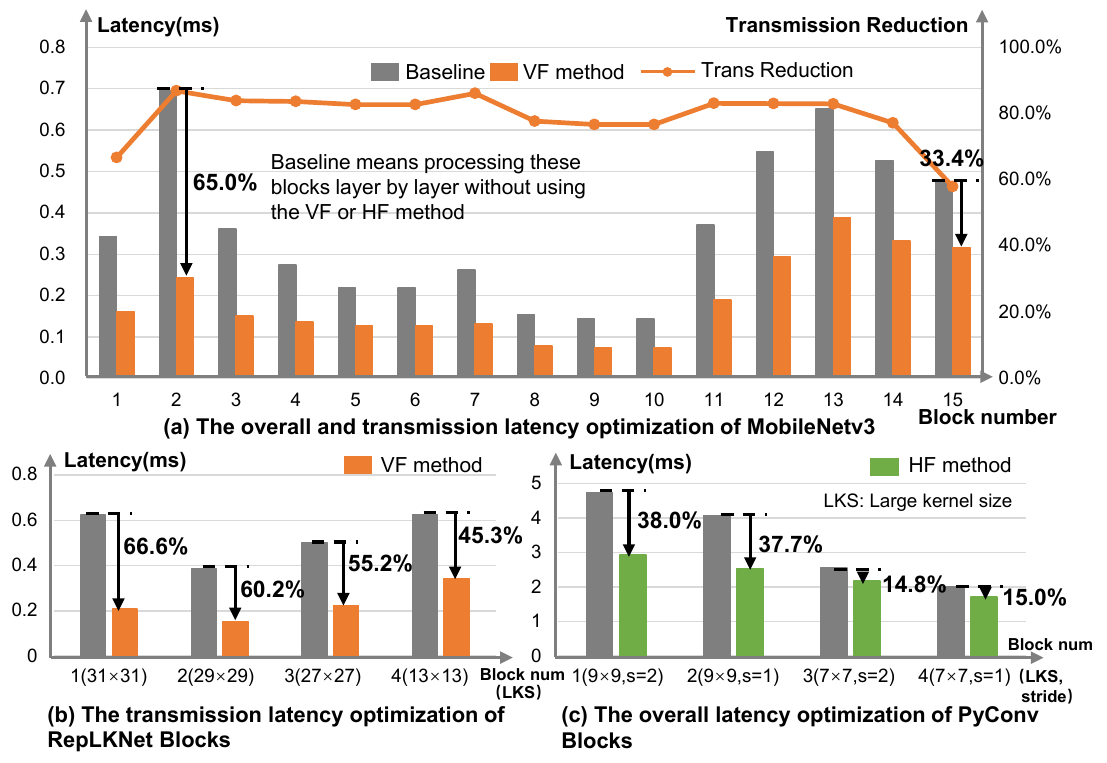}
  \vspace{-0.1in}
  \caption{Performance improvement (PI) of different typical blocks. (a) PI of MBconv blocks in Mobilenetv3 with VF method. (b) PI of RepLK blocks with VF method. (c) PI of PyConv blocks with HF method.
  }\label{Fig4}
 \vspace{-0.18in}
\end{figure}

\WX{The layer-level and block-level experimental results of the proposed methods are provided, with $Tiy=Niy$ and $Tif=Nif$ set to avoid optimization effects from the tiling strategy.}
Fig.~\ref{Fig3} (a) displays the results of performance comparisons between the proposed accelerator and the baseline evaluated on CONV operations with varying kernel sizes.
In the baseline, AKCV operations are performed by splitting large kernels into smaller ones (kernel size of 3$\times$3, 5$\times$5, or 7$\times$7), with the overlapped data stored in line buffers.
Our design, incorporating the proposed Z-flow method and Kseg scheme, demonstrates stable performance compared to the baseline.
\F{Besides, it significantly reduces the resource usage for data arrangement in the CONV engine as depicted in Fig.~\ref{Fig3} (b).}
Fig.~\ref{Fig4} \WX{shows the performance enhancements achieved through} the block-level optimization methods for some typical block structures.
Firstly, for MBconv blocks in MobileNetV3, the proposed VF method achieves significant reductions of up to 65\% in overall latency, saving approximately 80\% of the transmission latency among fused layers.
For RepLK blocks, the corresponding reductions in transmission latency range from 45.3\% to 66.6\%.
Additionally, the proposed HF method improves computational efficiency for PyConv blocks by optimizing both DRAM transactions and PE utilization, which is particularly efficient for block structures mismatched with the computing architecture.

\vspace{-0.08in}
\subsection{Comparison with Related Works}
\vspace{-0.02in}
Due to the lack of prior research on the deployment of emerging large-kernel CNNs, we choose several FPGA-based inference accelerators that implement CNN models with similar block structures, such as MobileNetV2 and ResNet-50, to facilitate fair comparisons.
Specifically, dedicated hardware architectures are proposed for light-weight CNNs in \cite{MobileNetV2_1} and \cite{MobileNetV2_3}, but their fixed computing engines designed for 3$\times$3 convolutions restrict efficient support for AKCV operations.
In\cite{2022TCASI}, the DWCV or pooling layer is directly performed after the CONV layer to reduce redundant memory accesses.
However, this complex computing dataflow arrangement makes it difficult to process the block structures with large-kernel operations.
Moreover, a reconfigurable row stationary-based CNN accelerator is proposed in \cite{2023TCASI}, optimized by a software-hardware co-design framework called Agamotto for the end-to-end execution of CNNs.
A unified architecture is developed in \cite{ResNet2} to efficiently process self-attention, multi-head attention, and convolution operations.
\WX{However, in these works, the computation parallelism is explored in the $Nif$ dimension, leading to inefficient PE utilization in certain layers (e.g., for the first layer, 3/32 utilization in \cite{ResNet2} versus 1 utilization in our design).}

Thanks to the proposed optimization methods, our design demonstrates outstanding performance, even for implementing CNNs with large kernels.
By exploring optimized loop tiling variables, the performance is further enhanced.
Specifically, as exhibited in Table \ref{table3}, the proposed accelerator achieves up to 3.25$\times$ and \F{3.91$\times$} better DSP efficiency when performing the inference of MobileNetV2 and ResNet-50, respectively, compared to state-of-the-art accelerators.

\vspace{-0.08in}
\section{Conclusion}
In this paper, a highly flexible and reconfigurable FPGA inference accelerator is proposed to efficiently support \WX{AKCV operations.}
Thanks to the proposed layer-level and block-level optimization methods, our design demonstrates excellent performance in processing CONV and DWCV layers, as well as typical block structures. 
The proposed accelerator is evaluated on Intel Arria 10 SoC FPGA for multiple prevalent CNNs, especially for RepLKNet-31 and PyConvResNet-50 with large kernels, \WX{achieving inference throughputs of 169.68 GOPS and 244.55 GOPS, respectively.}

\begin{table}[t]
\fontsize{5.8pt}{10.1pt}\selectfont
\caption{COMPARISON WITH OTHER WORKS}\label{table3}
\vspace{-0.25in}
\begin{center}
    \begin{tabular}{|c|c|c|c|c|c|c|c|}
        \hline
        \textbf{Work} & \textbf{\tabincell{c}{Freq.\\(MHz)}} & \textbf{Network} & \textbf{Platform} & \textbf{\tabincell{c}{Kernel\\size(max)}} &\textbf{DSP} & \textbf{\tabincell{c}{\tabincell{c}{TPUT \\(GOPS)}}} & \textbf{\tabincell{c}{DSP\\Eff.${^a}$}} \\
        \hline
        {\cite{MobileNetV2_1}} &150 & {MobileNetV2} & {Virtex-7} & {7$\times$7} &2160 & {181.8} & {0.56} \\
        \hline    
        {\cite{MobileNetV2_3}} &{200} & {MobileNetV2} & {\F{XCK325T}} & {7$\times$7} &{704} & {\F{197.1}}  & {\F{1.40}}\\ 
        \hline  
        {\cite{2022TCASI}} &{200} & {MobileNetV2} & {Arria 10} & {7$\times$7} &{607} & {188.2}  & {1.55}\\ 
        \hline
        {\cite{2023TCASI}} &200 & {ResNet-50} & {VCU118} & {7$\times$7} &{2048} & {287.2}  & {\F{0.70}} \\
        \hline
        {\cite{ResNet2}} &200 & {ResNet-50} & {XCVU37P} & {7$\times$7}  &{1260} & {590.0}  & {2.34} \\
        \hline 

        \multirow{4}{*}{\textbf{Ours}} & \multirow{4}{*}{200} & {MobileNetV2} & \multirow{4}{*}{Arria 10} & \multirow{4}{*}{\textbf{Arbitrary}}  &\multirow{4}{*}{522} & {\textbf{190.4}}  & {\textbf{1.82}} \\ 
        \cline{3-3} \cline{7-8} & & {RepLKNet-31} &  &  &  &{\textbf{169.6}} & {\textbf{1.62}}\\ 
        \cline{3-3} \cline{7-8} & & {ResNet-50} &  &  &  &{\textbf{286.2}} &  {\textbf{\F{2.74}}}\\ 
        \cline{3-3} \cline{7-8} & & {\tabincell{c}{PyConv-\\ResNet-50}} &  &  & & {\textbf{244.5}} & {\textbf{2.34}}\\
        \hline
    \end{tabular}
\begin{tablenotes}
        \item 
              $^{a}$ DSP Efficiency = (Throughput / DSP / Frequency)$\times 10^3$.\\
\end{tablenotes}
\label{tab1}
\end{center}
\vspace{-0.31in}
\end{table}
\vspace{-0.08in}
\section*{Acknowledgment}
\vspace{-0.02in}
This work was supported in part by the National Natural Science Foundation of China under Grant 62174084 and 62341408, the AI \& AI for Science Project of Nanjing University, and the National Key R\&D Program of China under Grant2022YFB4400604. 

\normalem
\bibliographystyle{IEEEtran}
\bibliography{lkacc}	

\end{document}